\documentstyle[aps,twocolumn,epsf]{revtex}

\begin{document}

\title{On the Phase Structure of the Micromaser System}

  \author  
  {Per Kristian Rekdal$^{1,}$\thanks{email: perr@phys.ntnu.no}
  and  Bo-Sture Skagerstam$^{1,2,}$\thanks{email: boskag@phys.ntnu.no}\\[3mm]
  {\small 
  $^{1}$Department of Physics,  The Norwegian University of Science and Technology,
  N-7034  Trondheim, Norway \\ 
$^{2}$Theoretical Physics Division, CERN, CH-1211 Geneva 23, Switzerland \\[-5mm]}}

\author{\begin{quotation}
\small We investigate, in an exact manner, 
       the phase structure of the micromaser 
       system in terms of the physical parameters at hand like  the atom cavity 
       transit time $\tau$, the atom-photon frequency detuning $\Delta \omega$, 
       the number of thermal photons $n_b$
       and the probability $a$ for a pump atom to be in its excited state.
       Phase diagrams are mapped out for
       various values of the physical parameters.
       At sufficiently large values of the detuning 
       we find a ``twinkling'' mode of the micromaser system.  A correlation length is used to 
       study 
       fluctuations close to the various phase transitions.
       \\[5mm]
\noindent PACS numbers 32.80.-t, 42.50.-p, 42.50.Ct
\end{quotation}
}

\maketitle

\newpage

\normalsize

There are few systems in physics which exhibit a rich struture  of
phase transitions that can be
investigated under
clean experimental conditions and  which  at the same time  can be studied by exact
theoretical methods. 
The micromaser system,  which is a remarkable  experimental realization of the idealized
system of a two-level  atom interacting with a second quantized
    single-mode electromagnetic  field, provides  us with such an example  
(for reviews and references see e.g. 
    \cite{Walther88}). 

Many features of the micromaser are of general interest.
It can e.g. be argued
that  the micromaser system is a simple
illustration of the conjectured  topological origin of second-order 
phase transitions\cite{Casetti99}.
Various aspects of
stochastic resonance can, furthermore, be studied in this system\cite{Buchleitner98}. 
The micromaser also
 illustrates a feature  of non-linear dynamical systems: turning on
randomness may led to an increased  signal to noise ratio\cite{noise94}.

   In a typical realization of the micromaser,  the pump atoms which enter the cavity 
    are assumed to be prepared in an incoherent
    mixture, i.e. the density matrix $\rho_A$ of the atoms is diagonal
with diagonal matrix elements $a$ and $b$ such that  $a+b=1$. 
In terms of the dimensionless atomic flux parameter $N=R/\gamma$, where $R$ is the rate 
injected atoms and $\gamma$  is the damping rate of the cavity, 
the stationary photon number
   probability distribution  is then
   well known  \cite{Filipowicz86} and is given by

\begin{equation} \label{p_n_eksakt}
  p_n = p_0 \prod_{m=1}^{n} 
  \frac{n_b \, m +  N a q_m}{(1+n_b) \, m + N b q_m}~~~.
\end{equation}

  \noindent
   Here $q_m \equiv q(m/N)$ and

\begin{equation} \label{q}
    q(x) = \frac{x}{x + \Delta^2}~
           \sin^2\left(\theta
           \sqrt{ x + \Delta^2 }~ \right)~~~,
\end{equation}

   \noindent
   where we have defined the dimensionless detuning 
 $\Delta = \Delta \omega/(2g\sqrt{N})$ and pump $\theta = g\tau \sqrt{N}$
   parameters.
   Furthermore,  $g$ is the single photon Rabi frequency at zero detuning
   of the Jaynes-Cummings (JC) model\cite{Jaynes63}.
   The overall constant $p_0$ is determined by $\sum _{n=0}^{\infty}p_n =1$.

   The theory as  developed in Refs.\cite{Filipowicz86,Guzman89}
   suggests the existence of various phase transitions
   in the large $N$ limit as the parameter $\theta$
   is increased.   A natural order parameter
   is then the average photon ``density'' $\langle x\rangle $, 
   where $\langle ~\rangle$ denotes an average  with
   respect to the distribution Eq.~(\ref{p_n_eksakt}) and $x=n/N$. 
   An exact treatment, in the large $N$ limit,
   of the micromaser
   phases structure and the corresponding critical fluctuations in
   terms of a conventional correlation length  has been given in
   Refs.\cite{ElmforsLS95}. Spontaneous jumps in  $\langle n\rangle /N$ 
   and  large correlation lengths close to micromaser phase transitions have
   actually been observed experimentally \cite{Walther88,Walther97}.
   Most of the theoretical and
   experimental studies have, however, been limited to the case $a=1$ and
   $\Delta =0$. It is the purpose of the present paper to study the
   phase structure of the micromaser system for general $a$ and
   $\Delta$ using methods which are exact in the large $N$ limit\cite{ElmforsLS95}. 
   As will be argued below, several new
   intriguing physical properties of the micromaser system are then unfolded.

   In order to obtain an appropriate
   expression for $p_n$ which describes the various  micromaser phases,
   we notice that
   the equilibrium distribution in Eq.~(\ref{p_n_eksakt}) can be re-written by
   using the Poisson summation technique \cite{Poisson}.
   The equilibrium distribution then takes the form

\begin{equation} \label{p_x}
   p(x) = p_0 \sqrt{\frac{w(x)}{w(0)}} ~ e^{-N \, V(x)}~~,
\end{equation} 

   \noindent where we have defined an effective potential 
   $V(x) = \sum_{k=-\infty}^{\infty}V_k(x)$ and 
\begin{equation} \label{V_k}
   V_k(x) = - \int _0^x d\nu  \, \ln[\, w(\nu)\,] \, \cos(2\pi N k \nu)
   ~~,
\end{equation}
\begin{equation} \label{w_x}
   w(x) = \frac{n_b \, x + a \, q(x)}{(1+n_b)\, x + b \, q(x)}~~.
\end{equation}
   \noindent 
   In the large $N$ limit Eq.~(\ref{p_x}) can be simplified  
   by making use of a  saddle-point approximation. 
   The saddle-points  are then determined
   by $V_0^{\prime}(x)=0$.
 If such non-trivial saddle-points exist we say that
   they describe maser phases. In a topological analysis of the second-order
   phase transitions\cite{Casetti99} of the micromaser system, 
   $V_0(x)$ will play the role of a Morse function.

   If the only global minimum of $V_0(x)$ corresponds to $x=0$, we expand the
   effective potential around the origin. The micromaser is then in a
   thermal phase and we obtain

\begin{equation} \label{p_n_geo}
 p_n = p_0
       \left (
       \frac{n_b+a\,\theta^{2}_{eff}}{1+n_b+b\,\theta^{2}_{eff}} \right)^n~~,
\end{equation}

   \noindent
   which is normalizable provided $\theta^{2}_{eff}\,(a-b) < 1$, where we
   have defined $\theta^{2}_{eff}=\sin^2(\theta\Delta)/\Delta^2$. 
   If $a < (1+\Delta^2)/2$ 
   the distribution Eq.~(\ref{p_n_geo}) is always valid. If $\Delta =
   0$ normalizability requires that $\theta^2\,(a-b) < 1$,
   i.e. $\theta$ must be sufficiently small if $a > 1/2$. 
   If $\Delta \neq 0$ the mean value of photons
   obtained from Eq.~(\ref{p_n_geo}) will be a periodic function of
   $\Delta\theta$ with a maximum $(n_b +
   a/\Delta^2)/(1+(b-a)/\Delta^2)$ for 
   $\Delta\theta = (n+1/2)\pi$, where $n=0,1,...$\,. The corresponding
   minimum is $n_b$ and occurs for $\Delta\theta = n\pi$. 
   In the thermal phase the micromaser can
   therefore be in a ``twinkling'' mode for $\Delta \neq 0$, i.e. the
   mean number of photons can exhibit periodic variations 
   as a function of $\theta$.

   Non-trivial saddle-points of the effective potential $V_0(x)$, 
   which  exist only if $a>(1+\Delta^2)/2$, 
   can be parametrically represented in the form \cite{ElmforsLS95}

\begin{eqnarray} \label{x_phi}
  &&  x + \Delta^2 = (a-b) \sin^2 \phi ~~, \nonumber \\
  &&  \theta = \frac{1}{\sqrt{a-b}} \frac{\phi}{|\sin \phi |}~~,
\end{eqnarray}

   \noindent
   with $\phi \geq \phi_0 \equiv \arcsin(|\Delta|/\sqrt{a-b})$.
   These saddle-points do not depend on $n_b$. 
   If, for a given $\theta$, there are several saddle-points, the actual
   maser phase is described by the saddle-point which corresponds to
   the global minimum of $V_0(x)$.

   The first extremum of $V_0(x)$ occurs at $\theta_0^* \equiv \theta(\phi_0)$.
   Critical points, where new extrema of $V_0(x)$ 
   appear, are determined by $V_0^{\prime \prime}(x)=0$, i.e. 
   non-trivial solutions of  $\tan \phi=\phi$.
   Correspondning to the positive solutions $\phi=\phi_k$, where
   $k=1,2,...\,$, we have the critical pump parameters  
   $\theta_k = \phi_k/(|\sin \phi_k| \sqrt{a-b})$.

\begin{figure}[htp]
\unitlength=0.5mm
\vspace{5mm}
\begin{picture}(160,140)(0,0)
\includegraphics{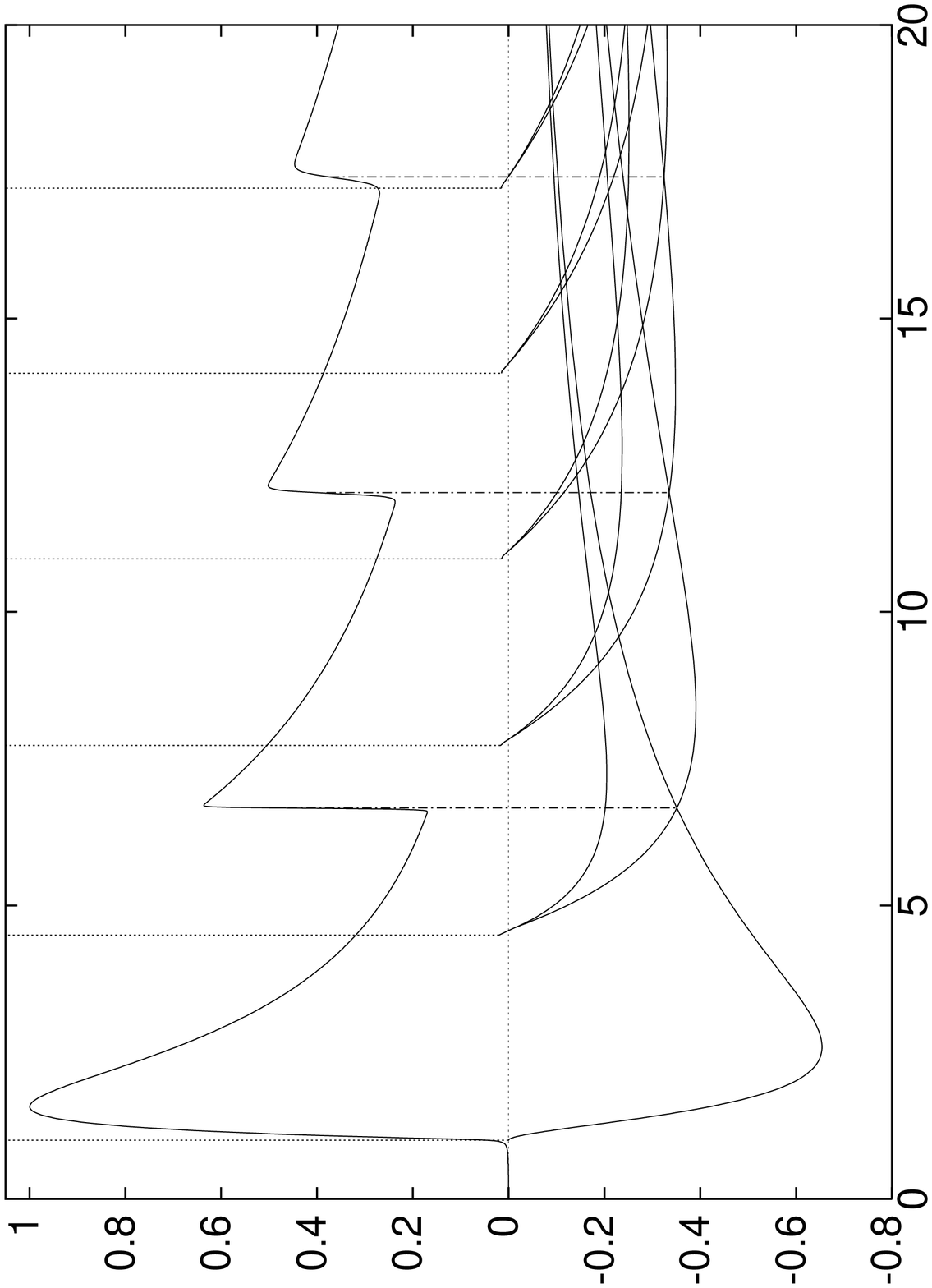}

   \put(25,130){\normalsize \boldmath$\frac{\langle n\rangle}{N}$}
   \put(25,52){\normalsize \boldmath$V_0(\theta)$}
   \put(18,145){\small  \boldmath$\theta_0^*$}
   \put(45,145){\small  \boldmath$\theta_1$}
   \put(70,145){\small \boldmath$\theta_2$}   
   \put(93.2,145){\small  \boldmath$\theta_3$}   
   \put(120,145){\small  \boldmath$\theta_4$}   
   \put(145,145){\small  \boldmath$\theta_5$}

   \put(58.5,52.5){\circle{5}}
   \put(46,40){\normalsize  $\theta^*_{01}\approx 6.66$}
   \put(101.3,53.5){\circle{5}}
   \put(86,32){\normalsize  $\theta^*_{12}\approx 12.03$}
   \put(144.3,54){\circle{5}}
   \put(128,40){\normalsize  $\theta^*_{23}\approx 17.41$}

   \put(54.5,-93.5){\vector(0,1){11}}
   \put(40,-101.5){\normalsize  $\theta_{t1}^*\approx 6.19$}

   \put(100.2,-93.5){\vector(0,1){11}}
   \put(86.5,-99.5){\normalsize  $\theta_{t2}^*\approx 11.91$}

\end{picture}
\vspace{-5mm}

\begin{picture}(160,140)(0,0)
\includegraphics{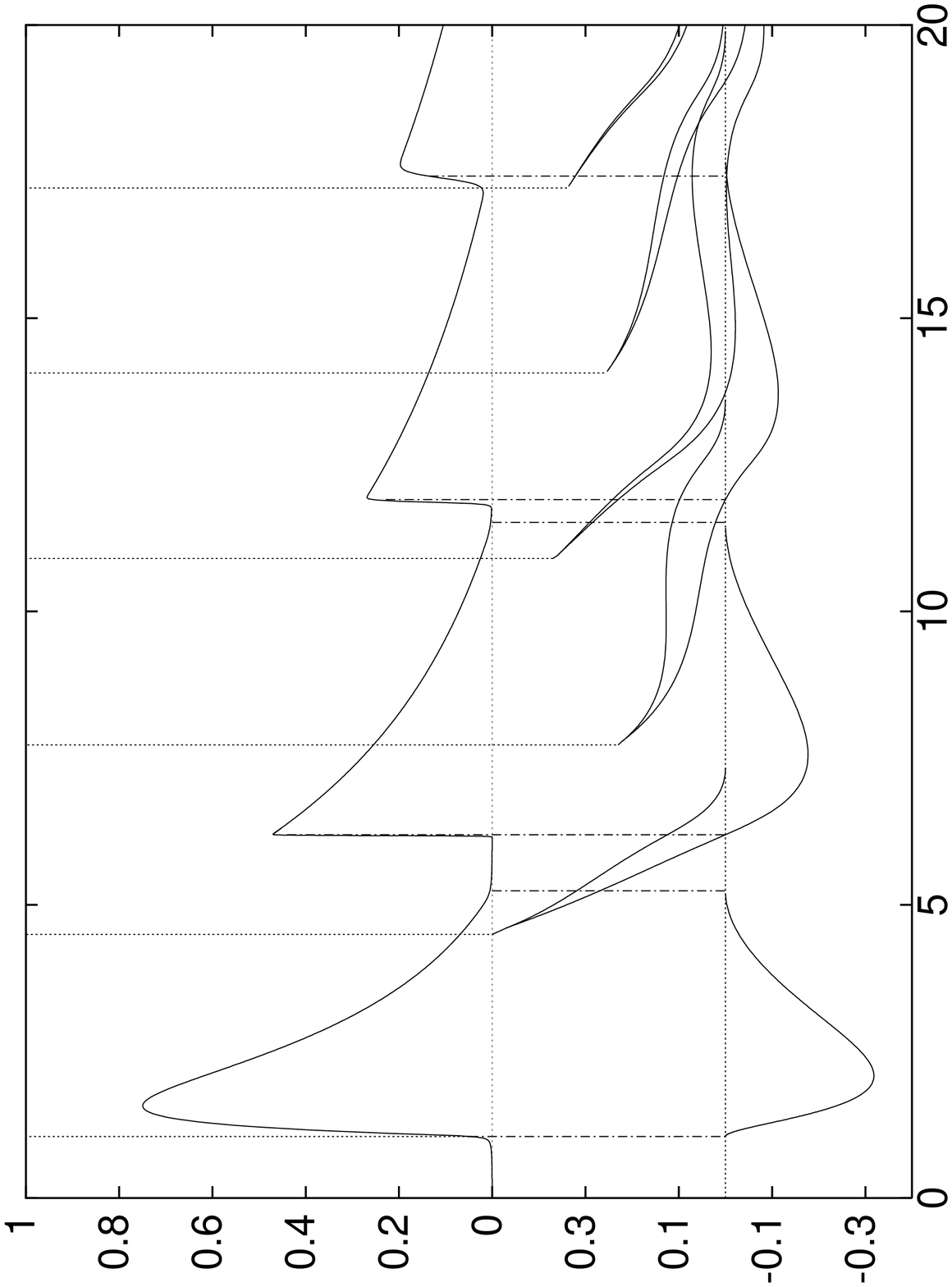}
       \put(79,10){\normalsize  \boldmath$\theta$}
       \put(144.4,48.3){\circle{5}}
       \put(129,34){\normalsize  $\theta^*_{23}\approx 17.43$}
       \put(25,55){\normalsize \boldmath$V_0(\theta)$}
       \put(25,130){\normalsize \boldmath$\frac{\langle n\rangle}{N}$}
\end{picture}
\caption[]{\protect\small  
           The extremal values of the effective potential 
           $V_0(\theta)$ when $a=1$, $n_b=0.15$, $\Delta =0$ (upper figure)
           and $|\Delta| =0.5$
           (lower figure). The critical parameters $\theta_0^*$ and $\theta_k$
           are shown as well as the pump parameters $\theta^*_{kk+1}$ 
           for co-existence of two maser phases.   
           $\theta_{k k+1}^*$, marked by a circle,  and $\theta_{tk}^*$ correspond to first-order 
           transitions.}
\label{V_theta_fig}
\end{figure}

 Using the substitution $\phi = \theta \sqrt{x+\Delta^2}$,
   the potential $V_0(x)$ in Eq.~(\ref{V_k})  can now be written in the form
\begin{eqnarray} \label{V_phi}
  V_0(\phi,\theta) = ~~~~~~~~~~~~~~~~~~~~~ && \nonumber \\  
  - \frac{2}{\theta^2} \, 
                \int_{\theta |\Delta|}^{\phi} d\phi \, \phi \, 
                \ln[\, \frac{n_b + a \, q(\phi,\theta)}{1+n_b + b\, 
                 q(\phi,\theta)}\,]~~,
\end{eqnarray}
   \noindent
   where $q(\phi,\theta) = \theta^2\sin^2\phi/\phi^2$.
    By choosing $\phi$ corresponding to Eq.~(\ref{x_phi}) the effective 
   potential is always at an extremum. We then get a multi-branched effective 
 potential $V_0=V_0(\theta)$.
 Branch $k$ is swept out when $\phi$ varies in the range $\phi_0 +k\pi
   \leq \phi \leq (k+1)\pi -\phi_0$, where $k=0,1,2,...$\,~.
   Except for the first branch ($k=0$), each of these branches is doubled-valued.
   One sub-branch then corresponds to a maximum 
   ($V_0^{\prime \prime}(x)<0$), which is swept out first as $\phi$ increases,
   and the other corresponds to a minimum 
   ($V_0^{\prime \prime}(x)>0$). 
   For a given branch $k$, these sub-branches coincide at the critical point
   $\theta_k$ (see e.g. Fig.~\ref{V_theta_fig}).

   In Fig.~\ref{V_theta_fig} we illustrate  such branches
   as well as the corresponding value of the order parameter
   $\langle x\rangle$.
   The transition from the thermal phase to the first maser
   phase at $\theta = \theta_0^*$  is always a  second-order phase transition. 
   The transition from one maser branch $k$
   to the neighboring maser branch $k+1$,  both with $V_0^{\prime \prime}(x)>0$,
   corresponds to a first-order phase transition and occurs at 
   $\theta = \theta^*_{kk+1}$. For various $k$ such
   critical points are marked by circles in
   Fig.~\ref{V_theta_fig}. As $\Delta$ increases the nature of these
   transitions changes. In Fig.~\ref{V_theta_fig} with $|\Delta| =0.5$ the first 
   two maser phases e.g. never intersect. Instead they intersect with the
   thermal phase. This phenomena corresponds to the twinkling behavior
   in the thermal phase even though we do not have a strictly
   periodic behavior. The twinkling phenomena will now, however, be more
   pronounced since, in the large $N$ limit, the maser will be
   ``dark'' in the thermal phase.
   For $a=1$ and $n_b = 0.15$ the phase separation of the first two maser
   phases occurs at $|\Delta| \approx 0.408$.
   With increasing value of $\Delta^2 \leq 1$ more phases are separated.

\begin{figure}[htp]
\unitlength=0.5mm
\begin{picture}(160,140)(0,0)
\includegraphics{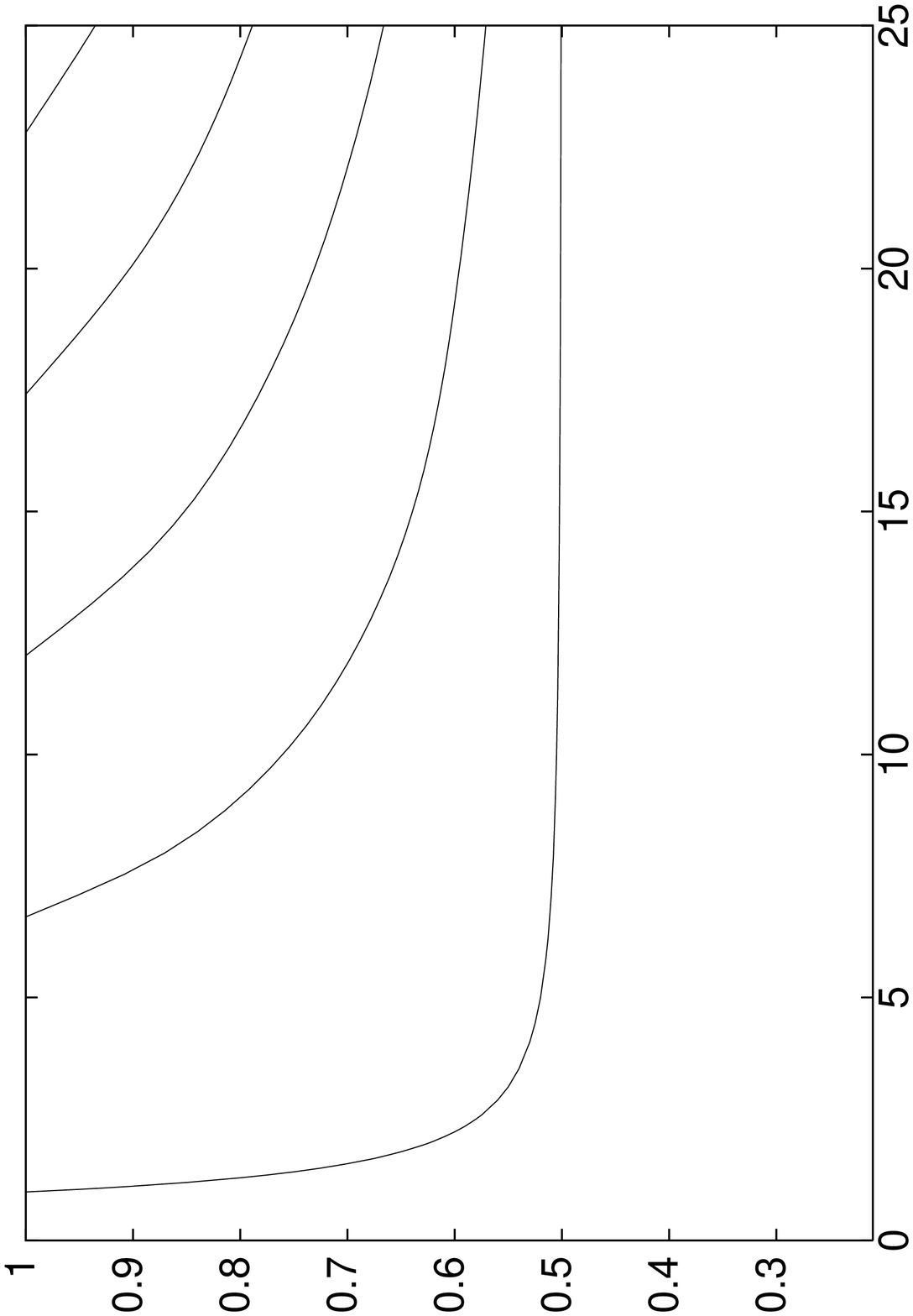}

   \put(-10,89.2){\normalsize  \boldmath$a$}
   \put(60,50){\normalsize  Thermal phase}
   
   \put(6,145){\normalsize  \boldmath$\theta_0^*$}
   \put(44,145){\normalsize  \boldmath$\theta_{01}^*$}
   \put(76,145){\normalsize  \boldmath$\theta_{12}^*$}
   \put(110,145){\normalsize  \boldmath$\theta_{23}^*$}   
   \put(146,145){\normalsize  \boldmath$\theta_{34}^*$}

   \put(15,50){\normalsize  $\Delta = 0$}
   \put(15,40){\normalsize  $ n_b = 0.15$}

\end{picture}
\vspace{-8mm}

\begin{picture}(160,140)(0,0)
  \put(81,10){\normalsize  \boldmath$\theta$}
  \put(-10,91.7){\normalsize  \boldmath$a$}
  \put(15,50){\normalsize  $|\Delta| = 0.5$}
  \put(15,40){\normalsize  $ n_b = 0.15$}
  \put(60,50){\normalsize  Thermal phase}
  \put(157,132){\circle{5}}
\put(117,136){\circle{5}}
\includegraphics{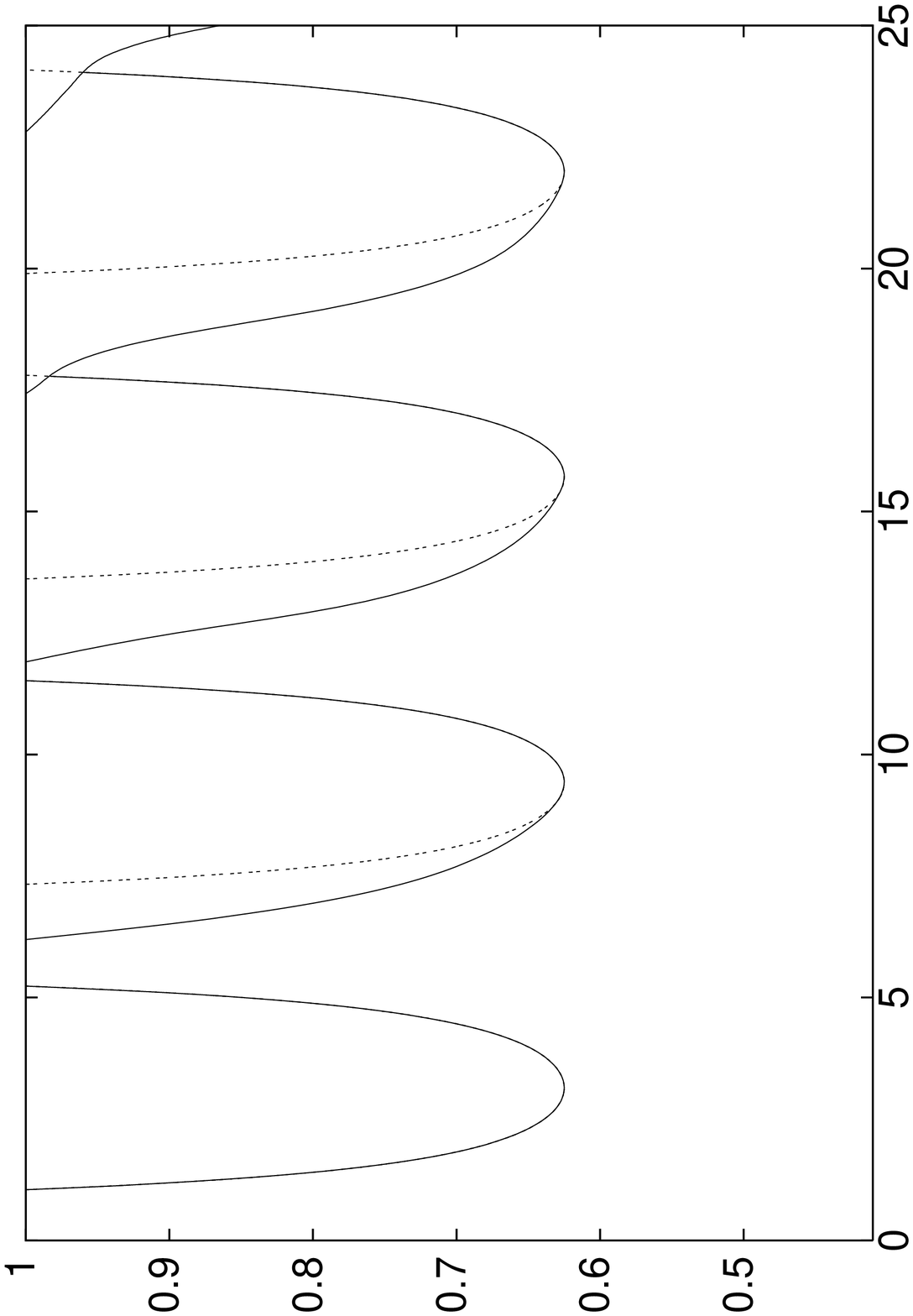}
\end{picture}
\caption[]{\protect\small 
           Phase diagrams for the same set of parameters as in Fig.~\ref{V_theta_fig}. 
           The upper figure is discussed in 
           the main text. The dashed curves in the lower figure correspond to the
           lines determined by $\theta^{2}_{eff}\,(a-b) = 1$. When not visible they
           overlap with the solid critical lines. The first critical line
           corresponds to a second-order (thermal-maser) transition. For any  
           other critical
           line the transition is first-order (second-order) to the left
           (right) of its  minimum (determined by $\sin^2(\Delta\theta)=1$)
           unless it intersects with another critical line. Triple points are
           indicated by circles.
}
\label{phase_fig} 
\end{figure}

   Critical
   transition lines can now be determined by considering 
   intersections between the various micromaser phases.
   Hence, we can
   construct a phase diagram in e.g. the $a$-$\theta$-parameter space. 
   The result is shown in Fig.~\ref{phase_fig} for the same parameters as
   in Fig.~\ref{V_theta_fig}. In the upper phase diagram in Fig.~\ref{phase_fig} the
   various micromaser phases are well separated, i.e. they 
   are uniquely separated by the critical lines.
   The first critical
   line is in general determined analytically by the condition 
   $\theta^{2}_{eff}\,(a-b) = 1$. 
   The other critical lines  are determined by means of analytical and 
   numerical methods.
   In the lower phase diagram of Fig.~\ref{phase_fig} with $|\Delta| = 0.5$, 
   micromaser phases which appear for sufficiently
   large values of  $\theta$ will again be well separated.

\begin{figure}[htp]
\unitlength=0.5mm
\vspace{5mm}
\begin{picture}(160,140)(0,0)
\includegraphics{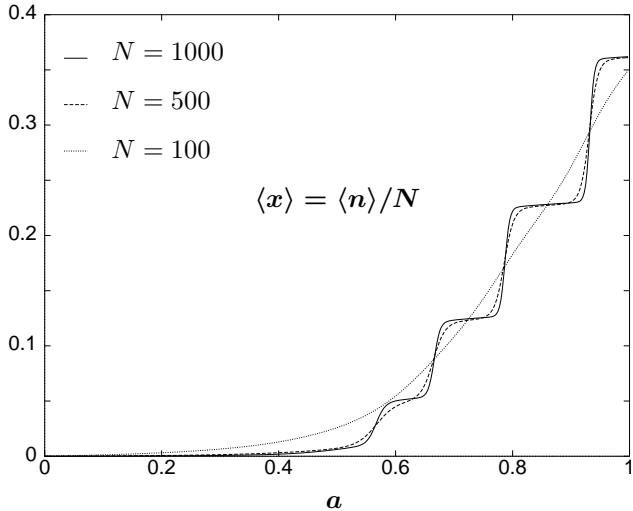}

   \put(79,10){\normalsize  \boldmath$a$}
   \put(60,90){\normalsize  \boldmath$\langle x \rangle = \langle n \rangle /N$}
   \put(22,129){\normalsize  $N=1000$}
   \put(22,116){\normalsize  $N=500$}
   \put(22,103){\normalsize  $N=100$}

\end{picture}
\caption[]{\protect\small 
           Plateaus of the order parameter $\langle x \rangle =
           \langle n \rangle /N$ 
           versus $a$ when $\Delta =0$, $n_b=0.15$ and $\theta =25$
           for various values of $N$. Each step corresponds to 
           a point on one of the transition lines in Fig.~\ref{phase_fig}.
\label{ordr_param_fig} }
\end{figure}

   The order parameter $\langle n \rangle/N$ can now also be studied as a function 
   of e.g. the probability $a$.
   Fig.~\ref{ordr_param_fig} shows such a plot. When we cross transition
   lines  in the phase diagram the order parameter will make discrete jumps 
   and hence also exhibits a plateau-like behavior.

   Let us now consider long-time correlations in the large $N$ limit
   as was first introduced in Refs.\cite{ElmforsLS95}.
   These correlations are most
   conveniently obtained by making use of the continuous time
   formulation of the micromaser system \cite{Lugiato87}. 
   The vector $p$ formed by the diagonal density matrix elements of the photon
   field then obeys the differential equation $dp/dt=-\gamma Lp$, where
   $L=L_C -N(M-1)$. Here $L_C$ describes the damping of the cavity , i.e.

\begin{eqnarray}
 && ( L_C)_{nm} = (n_b+1)[ \, n \delta_{n,m} - (n+1) \delta_{n+1,m} \, ] 
  \nonumber \\ && 
         ~~~~~~~~~~~   +  n_b[ \, (n+1)\delta_{n,m}
              - n \delta_{n,m+1} \, ] ~~,
\end{eqnarray}

   \noindent
   and $M=M(+) +M(-)$\, , where $M(+)_{nm} = b q_{n+1} \delta_{n+1,m} + 
   a(1-q_{n+1}) \delta_{n,m}$ and $M(-)_{nm} = a q_{n} \delta_{n,m+1} + b(1-q_{n}) 
   \delta_{n,m}$ have their origin in the JC-model \cite{Jaynes63,ElmforsLS95}.
   The lowest eigenvalue $\lambda_0=0$ of $L$ then determines the
   stationary equilibrium solution $p=p^0$ as given by
   Eq.~(\ref{p_n_eksakt}).
   The next non-zero eigenvalue $\lambda$ of $L$, which
   we  ~determine ~numerically, ~will then  ~determine typical 
scales for the approach to the stationary situation.
   The joint probability for observing
   two atoms, 
   with a time-delay $t$ between 
   them, can now be used in order to define a correlation length $\gamma^A(t)$\cite{ElmforsLS95}.
At large times $t\rightarrow \infty$, 
   we define the atomic beam correlation length $\xi_A$ by \cite{ElmforsLS95}

   \begin{equation}
     \gamma_A(t) \sim e^{-t/\xi_A}~~,
   \end{equation}

\vspace{5mm}
\begin{figure}[htp]
\unitlength=0.5mm
\vspace{5mm}
\begin{picture}(160,140)(0,0)
\includegraphics{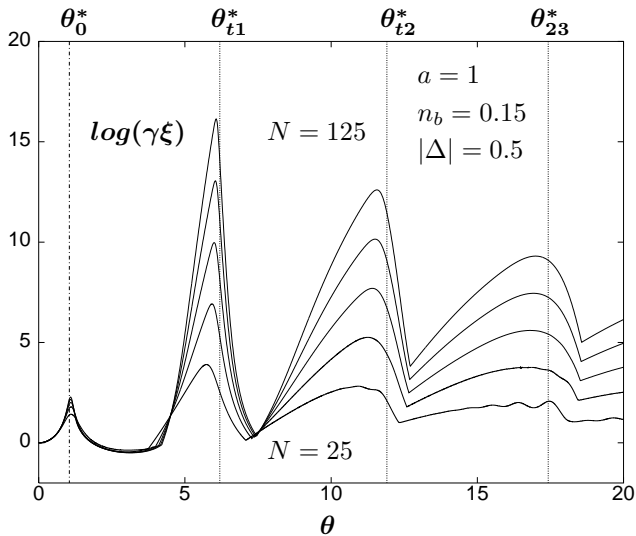}
   \put(79,10){\normalsize  \boldmath$\theta$}
   \put(18,115){\normalsize  \boldmath$log(\gamma \xi)$}
   \put(65,115){\normalsize  $N=125$}
   \put(65,30){\normalsize  $N=25$}
   \put(105,130){\normalsize  $a=1$}
   \put(105,120){\normalsize  $n_b = 0.15$}
   \put(105,110){\normalsize  $|\Delta| = 0.5$}

   \put(10,145){\normalsize  \boldmath$\theta_0^*$}
   \put(50,145){\normalsize  \boldmath$\theta_{t1}^*$}
   \put(95,145){\normalsize  \boldmath$\theta_{t2}^*$}
   \put(135,145){\normalsize  \boldmath$\theta_{23}^*$}

\end{picture}
\caption[]{\protect\small 
           The logarithm of the correlation length  $\gamma \xi$ as a function of 
           $\theta$ for various values of $N (25,50,...125)$.
           The vertical lines indicate the critical values of $\theta$ (see
           Fig.~\ref{V_theta_fig}).}
\label{corr_fig}
\end{figure}


   \noindent
   which then  is determined by $\lambda$, i.e. $\gamma\xi_A=1/\lambda$.
   For photons we define a similar correlation length $\xi_{C}$. It
   follows that the correlation lengths are identical, i.e. 
   $\xi_A=\xi_C\equiv \xi$\cite{ElmforsLS95}.
   The correlation length $\gamma \xi$ is shown in Fig.~\ref{corr_fig}
   for $|\Delta|=0.5$.
   In the large $N$ limit, the clear peaks in the correlation length
   occur at the critical pump parameters
   $\theta_0^*$, $\theta_{tk}^*$ and $\theta^*_{kk+1}$. 
 In the thermal phase we
   then have $\gamma\xi \simeq 1/(1+ (a-b)\theta^2_{eff})$, provided we
   are not to close to a critical line. When we move from the thermal phase to
   a maser phase the correlation length will increase. At the critical
   line $(a-b)\theta^2_{eff}=1$, the correlation length will behave as
   $(\gamma\xi)_{crit}\simeq (a-b)\sqrt{N/(a + (a-b)n_b)}/2$, apart from a
   weak dependence of the detuning parameter $\Delta$.

   \noindent 
   At the transition lines $\theta=\theta^{*}_{kk+1}$ the methods of
   \cite{ElmforsLS95} 
can be used to show that the correlation length depends
   exponentially on the combination $N(a+n_b (a-b))$. One can then also show that,
   as $\Delta^2$ approaches its maximal value $a-b$ for saddle-points to
   exist, $\gamma\xi\rightarrow 1$ exponentially fast. In view of the recent experiments
on trapping states in the stationary field configuration of 
the micromaser \cite{weidinger99}, we notice that 
high peaks occur in the atomic correlation length for such states\cite{ElmforsLS95}. 
The dependence of
these trapping state peaks on the physical parameters  at hand can be studied using the
methods presented here.

   In conclusion we have studied, in the large $N$ limit, the various 
   phase transitions and critical fluctuations of  the micromaser system 
   at non-zero detuning and for pump atoms prepared in a statistical
   mixture.
   We have revealed new novel features of the system as e.g. the 
   plateaus in the number of photons  $\langle n \rangle /N$ 
   as a function of $a$ as well as a
   new twinkling mode of the micromaser system for large detuning. This
   twinkling phenomena has a close resemblance with the observed revivals
   of the micromaser system \cite{Walther88}.

   B.-S.S. wish to thank H. Walther for discussions, comments on the manuscript
  and for providing a
   guide to the experimental work. 
   P.K.R. acknowledges support by the Research Council of Norway
   under the contract No. 118948/410.


\begin{thebibliography}{xx}


\bibitem{Walther88}
        H. Walther,
        Physica Scripta {\bf T23} (1988) 165;
        Phys. Rep. {\bf 219} (1992) 263;
        ``{\sl Experiments With Single Atoms in Cavities and Traps~}''
        in ``{\it Fundamental Problems in Quantum Theory~}'', Eds. D. M. Greenberger
        and A. Zeilinger, Ann. N.Y. Acad. Sci. {\bf 755} (1995) 133;
        Proc. Roy. Soc. {\bf A454} (1998) 431;
        Laser Physics {\bf 8} (1998) 1; Physica Scripta {\bf T76} (1998) 138.

\bibitem{Casetti99} L. Caiani, L. Casetti, C. Clementi  and M. Pettini, 
                    Phys. Rev. Lett. {\bf 79} (1997) 4361;
                    L. Casetti, E.G.D. Cohen and M. Pettini, Phys. Rev. Lett. {\bf 82}
                    (1999) 4160.

\bibitem{Buchleitner98} 
  A. Buchleitner and R.N. Mantegna, Phys. Rev. Lett. {\bf 80} (1998) 3932.

\bibitem{noise94} A. Maritan and J.R. Banavar, Phys. Rev. Lett. {\bf 72} (1994) 1451;
                  B.-S. Skagerstam in ``{\it Applied Field Theory~}'', 
                  Eds. Choonkye Lee, Hyunsoo Min and Q-Han Park
                  (Chungbum Publ. House, Seoul, 1999).


\bibitem{Filipowicz86}
        D. Filipowicz, J. Javanainen and P. Meystre, 
        Opt. Comm. {\bf 58} (1986) 327,
        Phys. Rev. {\bf A34} (1986) 3077.


\bibitem{Jaynes63}
        E.T. Jaynes and F.W. Cummings,
        Proc. IEEE {\bf 51} (1963)  89.

\bibitem{Guzman89}
        A.M. Guzman, P. Meystre and E. M. Wright,
        Phys. Rev. {\bf A40} (1989) 2471.

\bibitem{ElmforsLS95}
        {P.~Elmfors, B.~Lautrup and B.-S.~Skagerstam}, ``{\sl Correlations as
        a Handle on the Quantum State of the Micromaser~}'', CERN/TH 95-154
        (cond-mat/9506058); Physica Scripta {\bf 55} (1997) 724;
        Phys. Rev. {\bf A54} (1996) 5171.



\bibitem{Walther97} 
        O. Benson, G. Raithel and H. Walther,
        Phys. Rev. Lett.{\bf 72} (1994) 3506, {\it ibid.} {\bf 75} (1995) 3446,
        and ``{\sl Dynamics of the
        MicroMaser Field~}'' in ``{\sl Electron Theory and Quantum
        Electrodynamics: 100 Years Later~}'', Ed. J.P. Dowling (Plenum
        Press, New York, 1997).















\bibitem{Poisson} 
        See e.g. R. Courant and D. Hilbert, ``{\sl Methods
        of Mathematical Physics~}'' (Interscience, New York, 1953) and
        M. Fleischhauer and W.P. Schleich, 
        Phys. Rev. {\bf A47} (1993) 4258.




\bibitem{Lugiato87}
        L. Lugiato, M. Scully and H. Walther,
        Phys. Rev. {\bf A36} (1987)  740.



\bibitem{weidinger99} M.Weidinger, B.T.H. Varcoe, R. Heerlein and H. Walther,
Phys. Rev. Lett. {\bf 82} (1999) 3795.


\end{thebibliography}
\end{document}